\documentclass[journal=jacsat,manuscript=article]{achemso}

\usepackage[version=3]{mhchem} % Formula subscripts using \ce{}

\author{Ella M. Gale}
\altaffiliation{A response to R. S. Williams at Faraday Discussion}
\altaffiliation{University of Bristol, Bristol, UK}
\affiliation[Unknown University]
{University of Bristol, Bristol, UK}
\email{ella.gale@bristol.ac.uk}

\title[Response]
  {My, and others', spiking memristors are true memristors: a response to R.S. Williams' question at the New Memory Paradigms: Memristive Phenomena and Neuromorphic Applications Faraday Discussion}

\abbreviations{IR,NMR,UV}

\begin{document}


\begin{abstract}
At the Faraday Discussion, in the paper titled `Neuromorphic computation with spiking memristors: habituation, experimental instantiation of logic gates and a novel sequence-sensitive perceptron model' it was demonstrated that a large amount of computation could be done in a sequential way using memristor current spikes (d.c. response). As these spikes are found in many memristors (possibly all), this novel approach could be highly useful for fast and reproducible memristor circuits. However, questions were raised as to whether these spikes were actually due to memristance or merely capacitance in the circuit. In this longer version of the Faraday Discussion response, as much information as is available from both published and unpublished data from my lab is marshalled together. We find that the devices are likely imperfect memristors with some capacitance, and that the spikes are related to the frequency effect seen in memristor hysteresis curves, thus are an integral part of memristance.
\end{abstract}

\section{Introduction}

At the New Memory Paradigms: Memristive Phenomena and Neuromorphic Applications Faraday discussion on my paper\cite{Fara} the question of whether or not the devices that I have been characterizing and using as memristors were, or were not, memristors came up. For years, I have been suggesting that the spikes I first observed with the devices in my lab 2012\cite{SpC0} were a critical part of memristance, and, in fact, that these current spikes were in fact the d.c. response of the memristor. This implies that memristors are capable of operating as a d.c. device, something about which there had been doubt previously. As it is not only with my memristors that one can observe this phenomenon, and, in fact others have seen similar spikes, figuring out whether the spikes are largely to do with memristance or largely to do with capacitance is critical. In this short letter, I will gather as much information that I have to hand to try and answer this question, and I will use R.S. William's question from the Faraday discussion, as he phrased the issue well. 

\section{Comment}

For the Faraday discussion, R. S. Williams wrote:
\begin{quotation}
`I'm afraid that the devices you show in your paper are {\bf not memristors at all, but are in fact leaky capacitors}. Just putting some form of titanium dioxide between two electrodes does not meant that you will obtain memristive behaviour if there is some type of ionic species, e.g. {\bf O vacancies, that acts as a mobile electronic dopant, or some type of phase change. You have neither in your system}, as can be seen in Fig. 1 of your paper\cite{Fara}, which shows a classical transient response of a capacitor after abruptly turning on a voltage. Sometimes one can see a similar transient in a circuit containing a dielectric between two electrodes - in that case, one needs to construct a model for the physical device that has a memristor in parallel with a capacitor to model it correctly. There is an effective parallel resistance in your system, either from leakage through capacitor or {\bf from the measurement circuit}. I have taken for Fig 1. and show that you can compute the RC time constant from the 1/e decay time of your capacitor, which is about 5 seconds (there appears to be a second and somewhat longer time scale in your system, which probably comes from electrons escaping from charge traps in your sol-gel oxide). You can then calculate the resistance of the effective parallel resistor from R=V/I, {\bf where V is the DC bias voltage (0.1 V? - it was not clear to me what the applied voltage was for Fig. 1)} and I is the asymptotic current ($\sim$2.7nA). If you guess for the voltage is correct, then I compute that the parallel resistance is ~37M$\omega$ and thus the capacitance would be $\sim$135nF. This is a fairly large number, {\bf but I did not see the size of the devices} in the paper so it is at least plausible. In principle, you can now build a nearly identical equivalent circuit from a set of standard linear capacitors and resistors, with no memristors involved.
\end{quotation}

R.S. Williams raises an interesting point: is this observed spike response of a memristor to a voltage step merely explained by a capacitor and resistor in parallel (as he suggests) or is it all explained by memristance, or is it explained by a memristor and capacitor in parallel? As my  memristor devices are physical devices, and not ideal circuit elements, and as they are made of a-TiO$_2$\cite{M0} spin-coated onto sputtered electrodes, they are highly unlikely to be modelled by pure memristor theories. So the question really is, are these devices (and thus the observed behaviour) due to a capacitor-resistor (RC) circuit, or a memristor-capacitor (MC) circuit.  

Since this is such an important point, I will attempt to marshal as much evidence as I have to elucidate which of these possibilities is the correct one. 

%\begin{figure}
%\centering
%\includegraphics[width=0.8*\textwidth]{Fig1.png}
%  \caption{Dr. Williams' slide, a reproduction of figure 1 from\cite{Fara} with equations written on.}
%  \label{fig.1}
%\end{figure}

\section{Response}

\paragraph{Point 1. Some of my devices are capacitors:}

R. S. Williams is correct that it is possible to make a capacitor with my set-up, and a capacitor V-I curve can be recorded, 3 were seen out of 64 devices\cite{M1}. An example V-I curve is shown below and is taken from \cite{M1}, fig 6d.

\begin{figure}
\centering
\includegraphics[width=\textwidth]{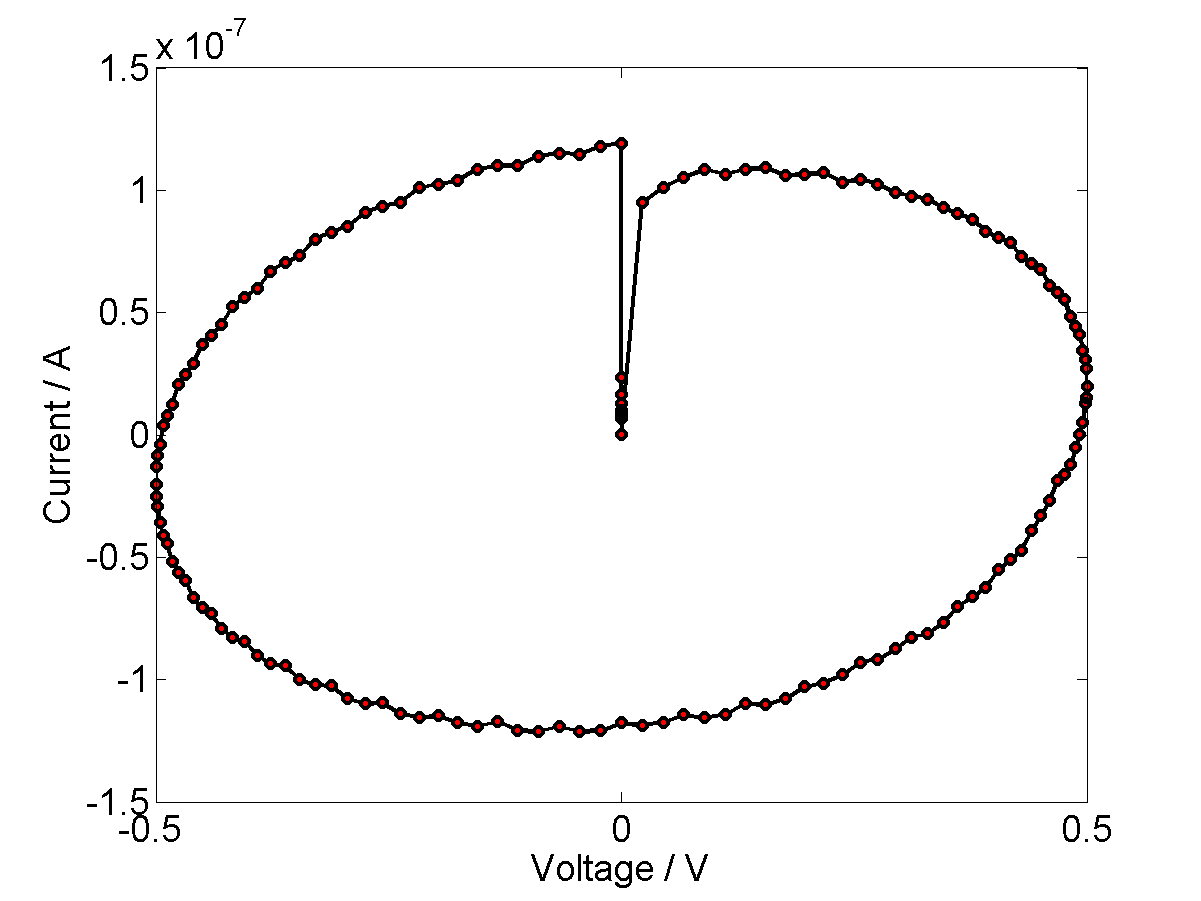}
  \caption{Capacitor V-I curve seen in one of my devices}
  \label{fig.2}
\end{figure}

\paragraph{Aside 1. The size of my devices:}
R. S. Williams asked about the size of my devices, they are relatively large. I should point out that at the time I was making these devices (2010), very few people had memristors of any type, so the ones I made, although not the smallest or best, were highly useful for testing out ideas, as that point, I believe the only people making and testing memristors in the UK was myself and T. Prodromakis. My devices had electrodes of width 1-6mm, with most of them being 5mm wide, the electrodes were sputtered aluminium onto plastic, Ti(OH)$_4$ solution was then spin-coated on and the devices left to dry either in clean room air (the work was done by Hewlett Packard in their clean room) or under vacuum before a second set of electrodes were sputtered on at 90 degrees to the first. The TiO$_2$ layer was 44nm thick, and often appeared slightly pinkish in colour. There are many more details in \cite{M1}.

\begin{figure}
\centering
\includegraphics[width=\textwidth]{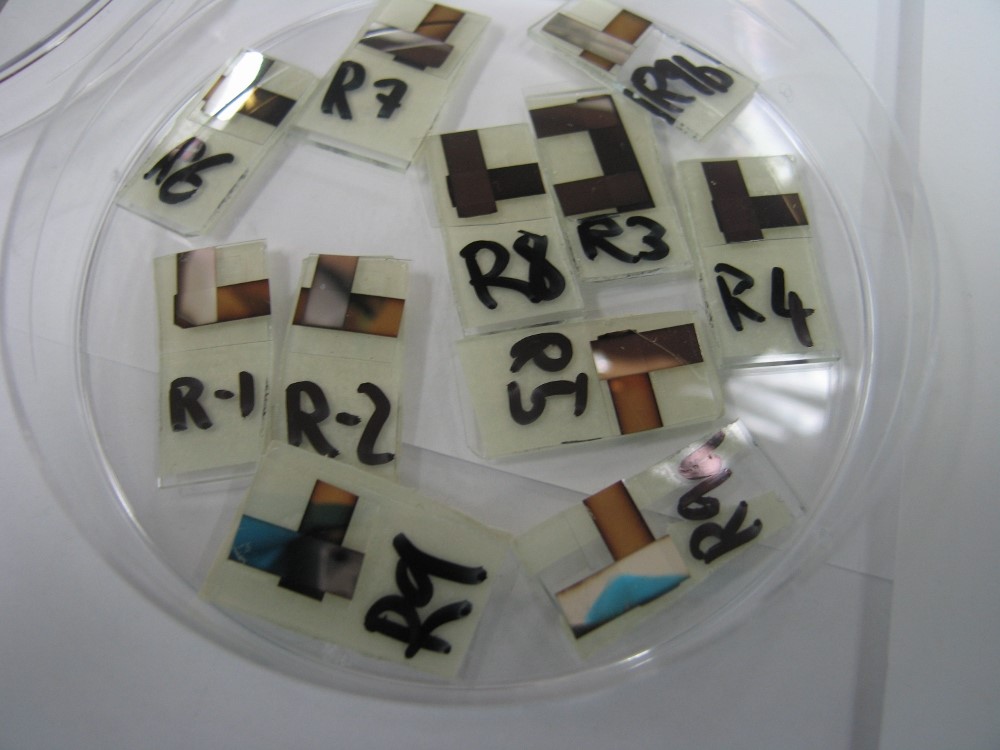}
  \caption{Memristor devices used in this work, shown in a petri dish for scale. These devices were sputtered onto plastic, but mounted on glass slides for testing}
  \label{fig.3}
\end{figure}

\paragraph{Aside 2: There are there Oxygen vacancy dopants in my system:}

R. S. Williams said `\emph{Just putting some form of titanium dioxide between two electrodes does not meant that you will obtain memristive behaviour if there is some type of ionic species, e.g. {\bf O vacancies, that acts as a mobile electronic dopant, or some type of phase change}. You have neither in your system},' and I am slightly confused by this point. I do have oxygen vacancies in my system, and given that they were sputter coated onto aluminium, I do not think I needed to dope the material in order to get them, the methodology itself introduced vacancies in the amorphous TiO$_2$. It’s worth pointing out that my memristors are based on Gergel-Hackett’s work\cite{GH}. There might well be some type of phase change, in my work I put the `curved' type behaviour down to a phase change mechanism, but elucidating how memristors/ReRAM work is difficult and not finished\cite{Review}. I also demonstrated in\cite{M1} that the aluminium electrodes were an essential part of getting the devices to work as memristors, I believe because the Al electrode is capable of making Al$_2$O$_3$ and thus acting as a source/sink of oxygen ions (sputtered Au electrode devices did not work), which does suggest that I have oxygen vacancies. My devices were always run the same way round, so the hysteresis was bigger in the positive V part of the curve, I believe this is because the Al electrode that is has Ti(OH)$_4$ sputtered on top of it is a different structure (perhaps a larger Al$_2$O$_3$ layer) than the electrode that was sputtered onto the TiO$_2$(gel) layer.  

\paragraph{Aside 3: My measurement kit has no leakage}
R. S. Williams wrote: `\emph{There is an effective parallel resistance in your system, either from leakage through capacitor or {\bf from the measurement circuit}.}'
There is no leakage capacitance from the measurement circuit as I used a Keithley electrometer set-up which was rated to very high accuracy and which is designed to test devices, so of course, it will not have capacitance leakage.

\paragraph{Point 2. Many of my devices exhibit the pinched hysteresis loop typical of memristors:}

\begin{figure}
\centering
\includegraphics[width=\textwidth]{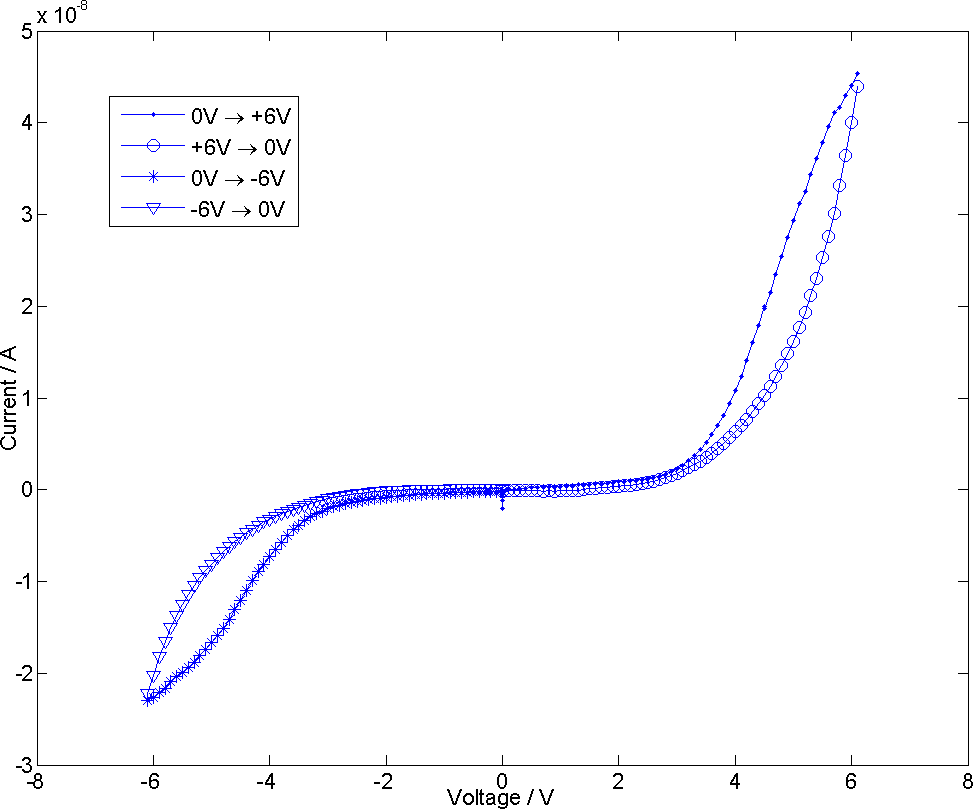}
  \caption{Example V-I curve, showing the clockwise switching of a device, note that this is for the much thicker hand assembled devices, not the thinner spin coated ones I published in my spiking research, although I found much the same behaviour.}
  \label{fig.4}
\end{figure}

\begin{figure}
\centering
\includegraphics[]{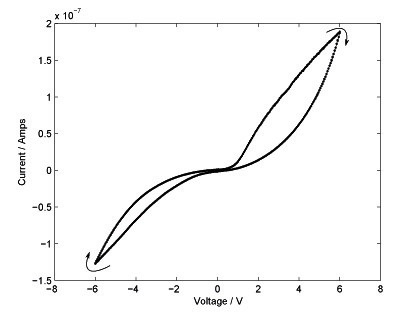}
  \caption{A pinched I-V curve from an Al-TiO$_2$-Al memristor, taken from\cite{M0}. Note that the pinched hysteresis curve is pinched at zero but does not cross at zero }
  \label{fig.5}
\end{figure}

\begin{figure}
\centering
\includegraphics[width=\textwidth]{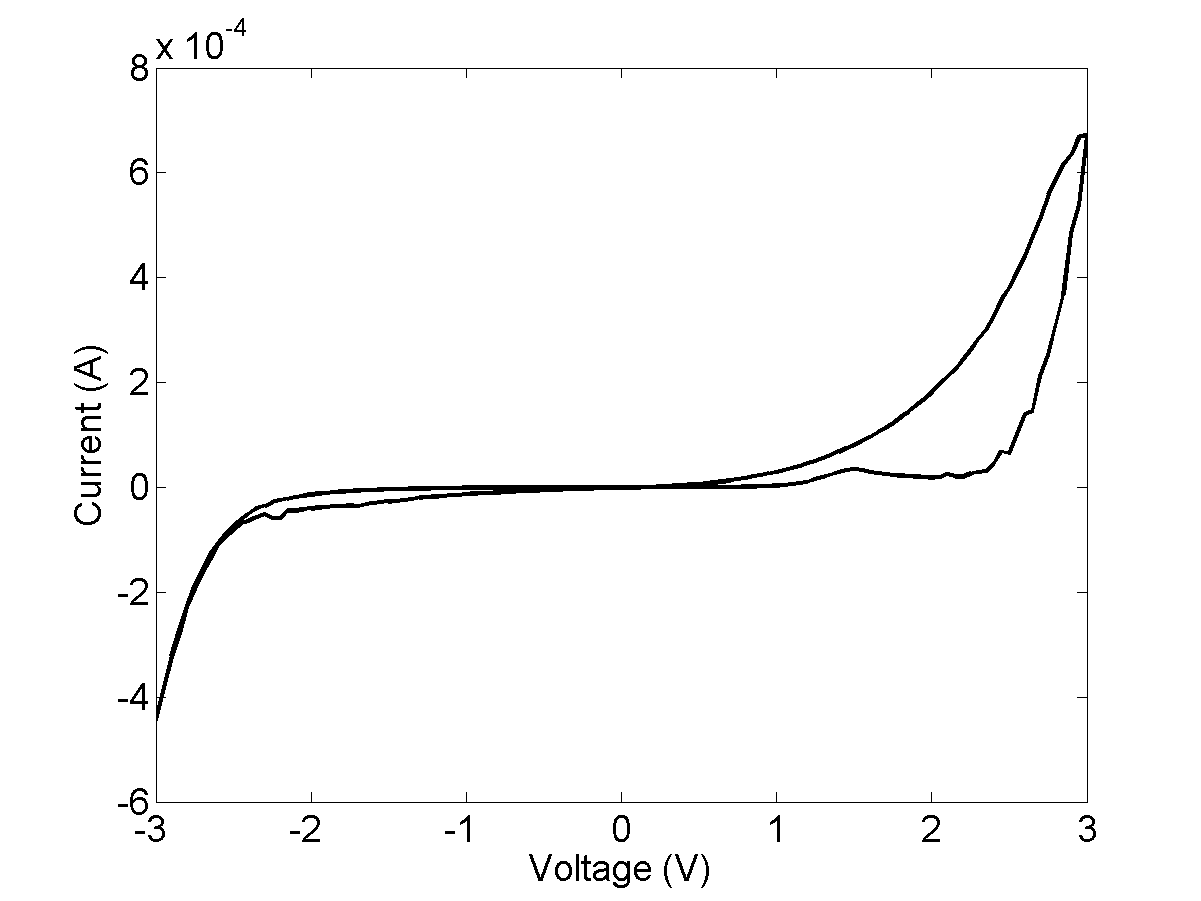}
  \caption{A `curved' memristor (taken from\cite{M1}}
  \label{fig.6}
\end{figure}

\begin{figure}
\centering
\includegraphics[width=\textwidth]{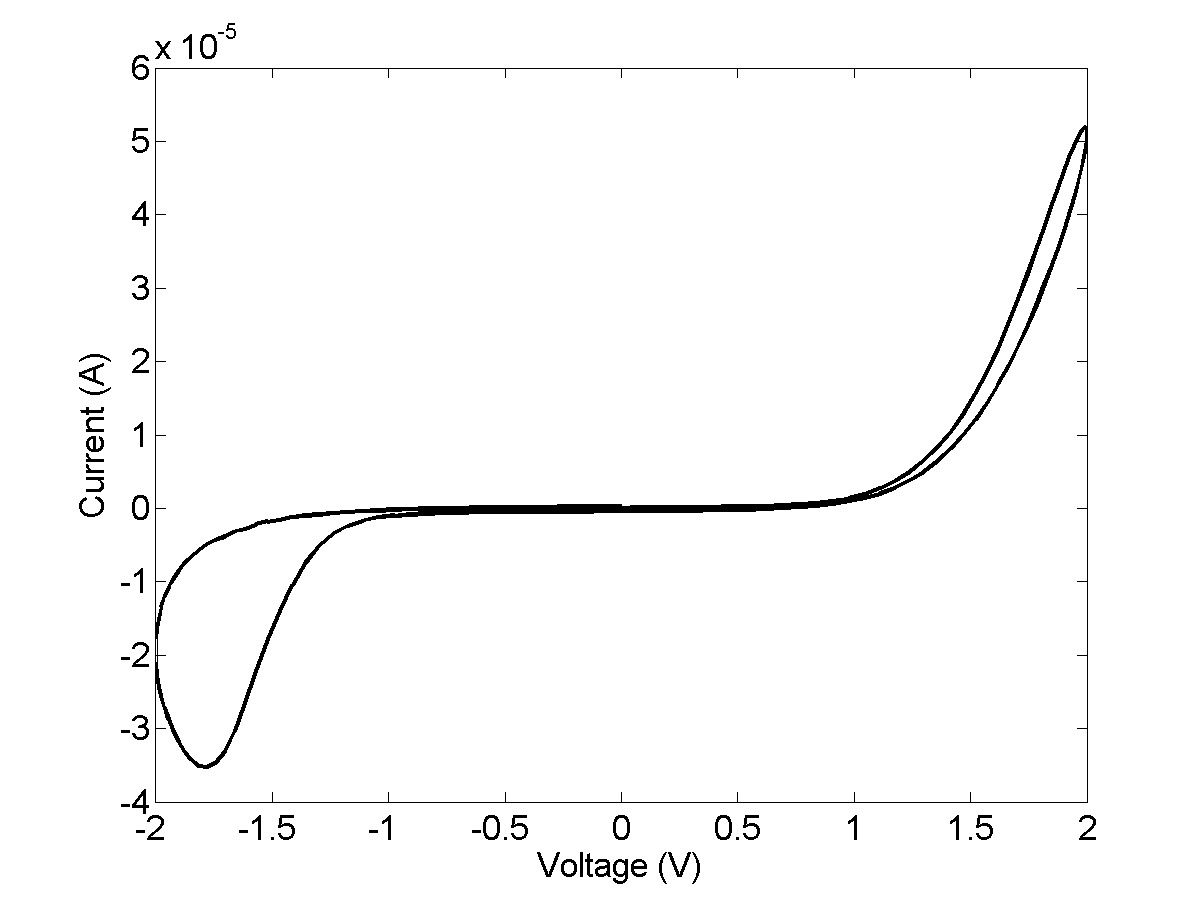}
  \caption{Another  `curved' type memristor (taken from \cite{M1})}
  \label{fig.7}
\end{figure}
\begin{figure}
\centering
\includegraphics[width=\textwidth]{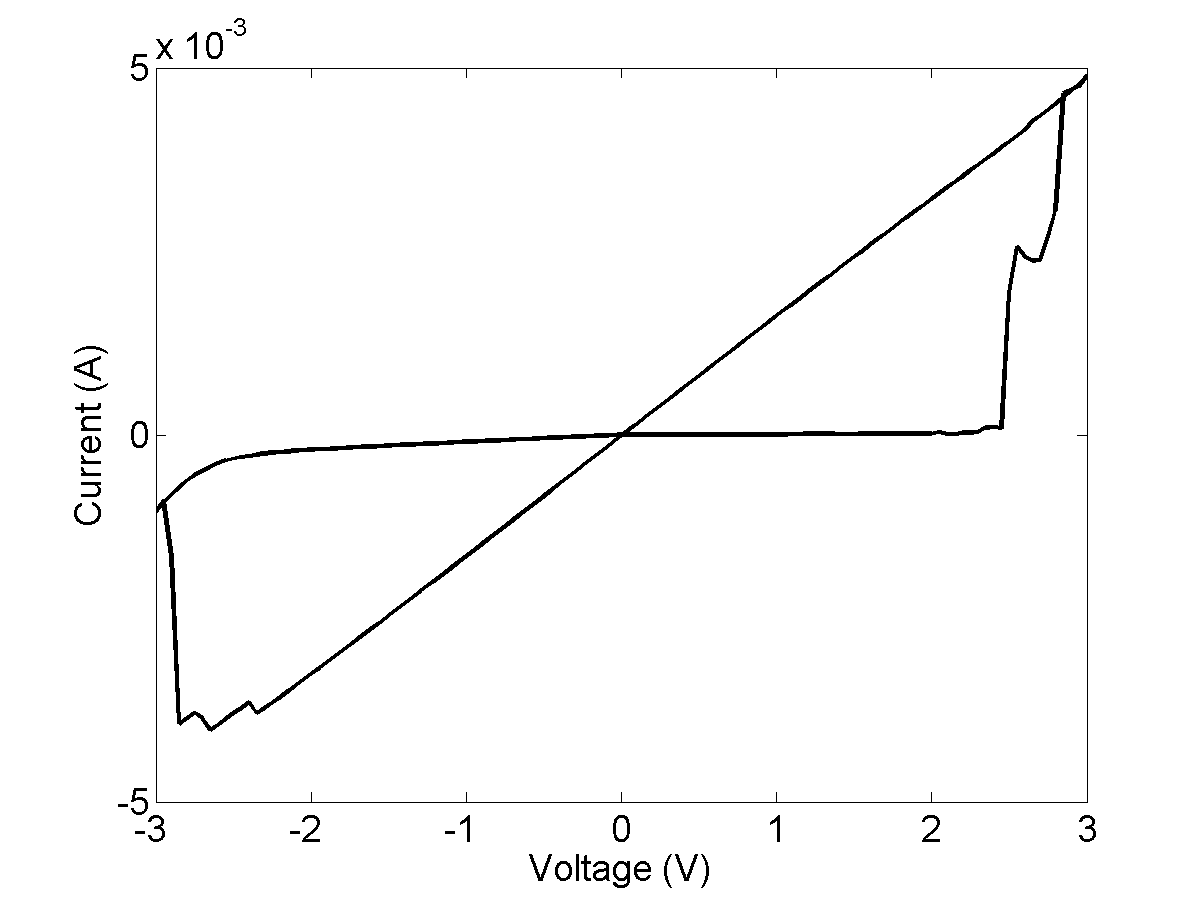}
  \caption{A `triangular' type memristor (taken from \cite{M1})}
  \label{fig.8}
\end{figure}

As my devices (like, I suspect, many other peoples’) were so variable in quality and behaviour, I developed a system of classifying their behaviour over a very small voltage range, so I could sort them into different types of memristors without accidentally forming them.

I classified the behaviour of the devices that I had made into various responses seen on a low voltage input (so as to avoid forming the devices). Those which displayed what I called Ohmic behaviour (i.e .they were a resistor over the range) then displayed `triangular' behaviour over a larger range (which at the time we called UPS or ReRAM-like behaviour, as at the point the ReRAM community tended to measure their devices over a positive range and see something they called (unipolar switching) UPS, this was also the point where the memristor and ReRAM communities were separate and had not come together around the idea that memristors and ReRAM were the same thing.) I suspect what I called `triangular' behaviour was due to a filamentary fuse-antifuse type mechanism, but actually figuring out the mechanism behind all of this was rather difficult and time consuming (and still not settled in the field, see \cite{Review}). I called devices that did triangular switching `type B' behaviour. These devices tended to have a resistance change of an order of magnitude or more, and never switched at the same point on repeated runs (there was a big problem of reproducibility), which was (and still is, I believe) shared by many other workers in the field.

Curved, or type A behaviour, was what I called the devices that exhibited more `memristor' type curves (remembering that at the time, memristor scientists were trying to reproduce L. O. Chua’s 'figure of 8' curves\cite{Chua1971} and ReRAM scientists were doing triangular V-I curves which switched between resistance states), these were pinched hysteresis loops, and they tended to be more reproducible and more likely to stay within the same current magnitude over multiple V-I runs. 

\paragraph{Point 3. My curved-type memristors do indeed have some capacitance, and these devices produce pinched V-I curves over a larger voltage range}

By classifying my devices at low voltage, I saw the following V-I curves, which I named `jelly bean' open curves for the shape. 

\begin{figure}
\centering
\includegraphics[width=\textwidth]{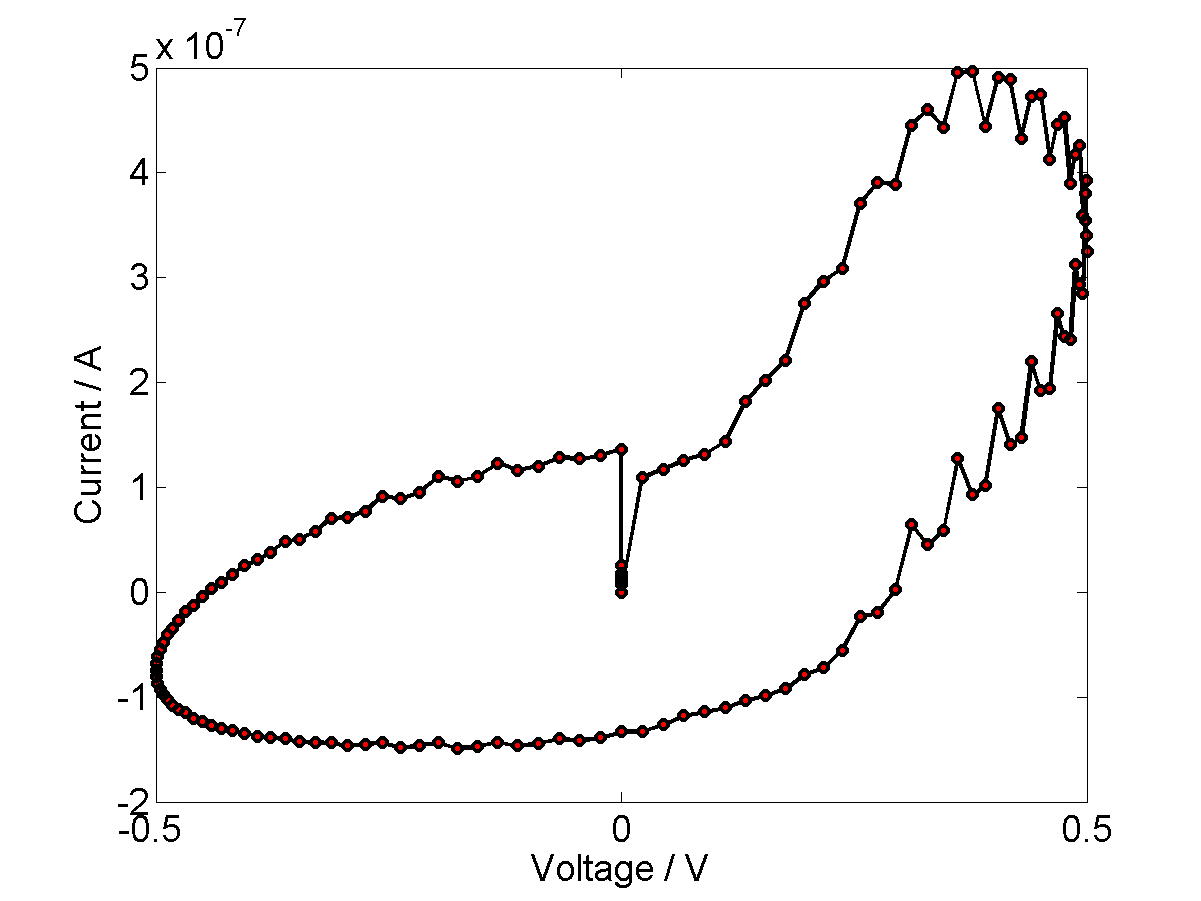}
  \caption{A 'curved' memristor (taken from \cite{M1}}
  \label{fig.8b}
\end{figure}

As these devices do not cross zero, there is obviously some capacitance in the devices. However, these open curves are not circles, so this is not a pure capacitor. (And my measurement system was a Keithley electrometer, so there is no source of capacitance elsewhere in the system). These open curves exhibit the same asymmetry I see in my devices' over a larger voltage range (namely the positive polarity voltage has a larger magnitude current response\footnote{Note that, I always test my devices the same way round}). And, the crucial point: {\bf these jelly bean open curve devices exhibit the pinched hysteresis loops (curved type switching) I showed above, and which are widely agreed to be indicative memristance}. This suggests to me that my devices are memristors that also possess a little capacitance which prevents them from crossing at 0, but which does not explain the memristance effects seen at higher voltages. 

\paragraph{Point 4: Part of the `Jelly bean' low-voltage V-I curves can be explained by a memristor theory, but seems to be missing a capacitance}

In order to do a test of several different memristor theories, I made some memristors of different electrode sizes, and characterised their hysteresis at low voltage. The hysteresis increases with electrode size. 

\begin{figure}
\centering
\includegraphics[width=\textwidth]{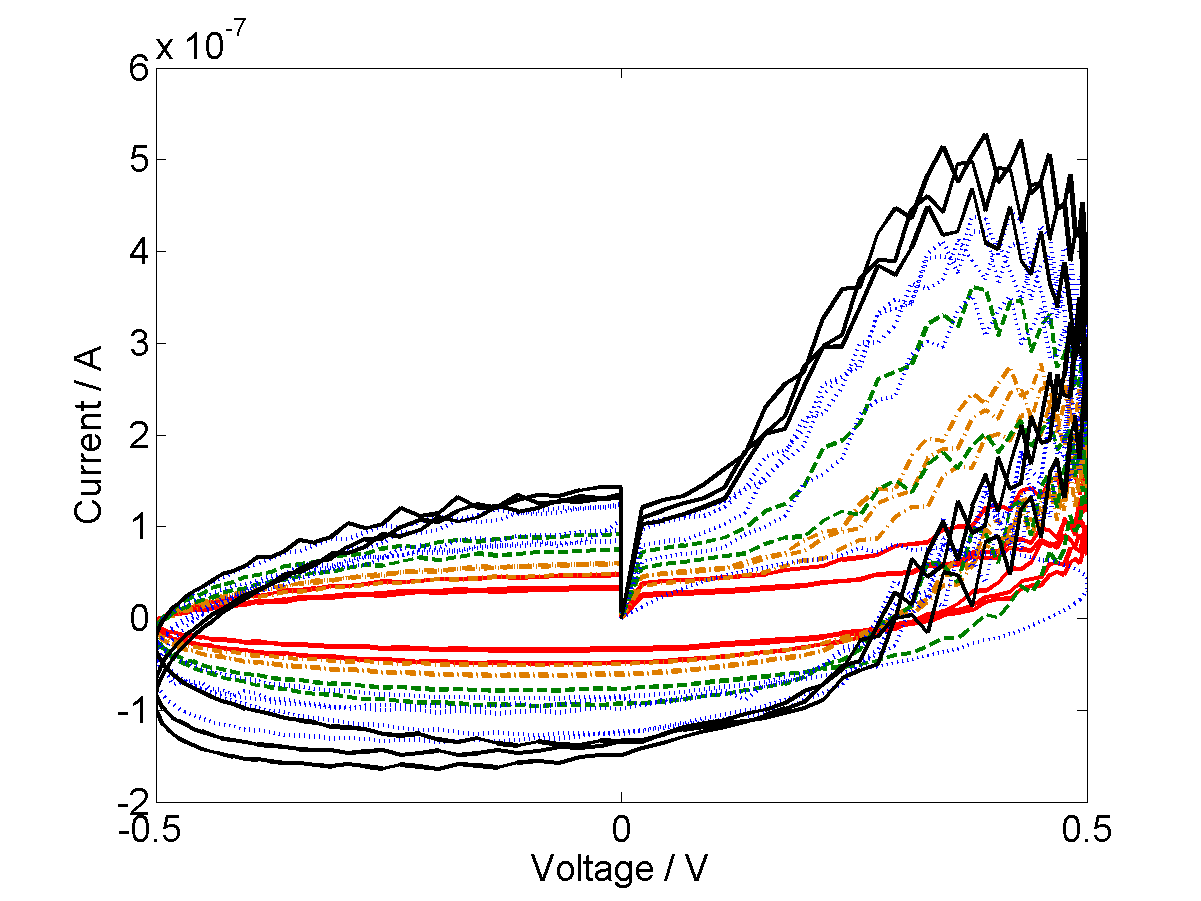}
  \caption{Recorded jelly-bean V-I curves seen for different electrode widths: Black: 5mm; blue dotted: 4mm; green-dashed 3mm; orange dot-dashed 2mm; red: 1mm. Data from actual device experiments.}
  \label{fig.9}
\end{figure}

Using the memory-conservation theory\cite{F0}, with the expected electrode widths, we can predict the V-I curves which would be expected in ideal memristors with these characteristics.

\begin{figure}
\centering
\includegraphics[width=\textwidth]{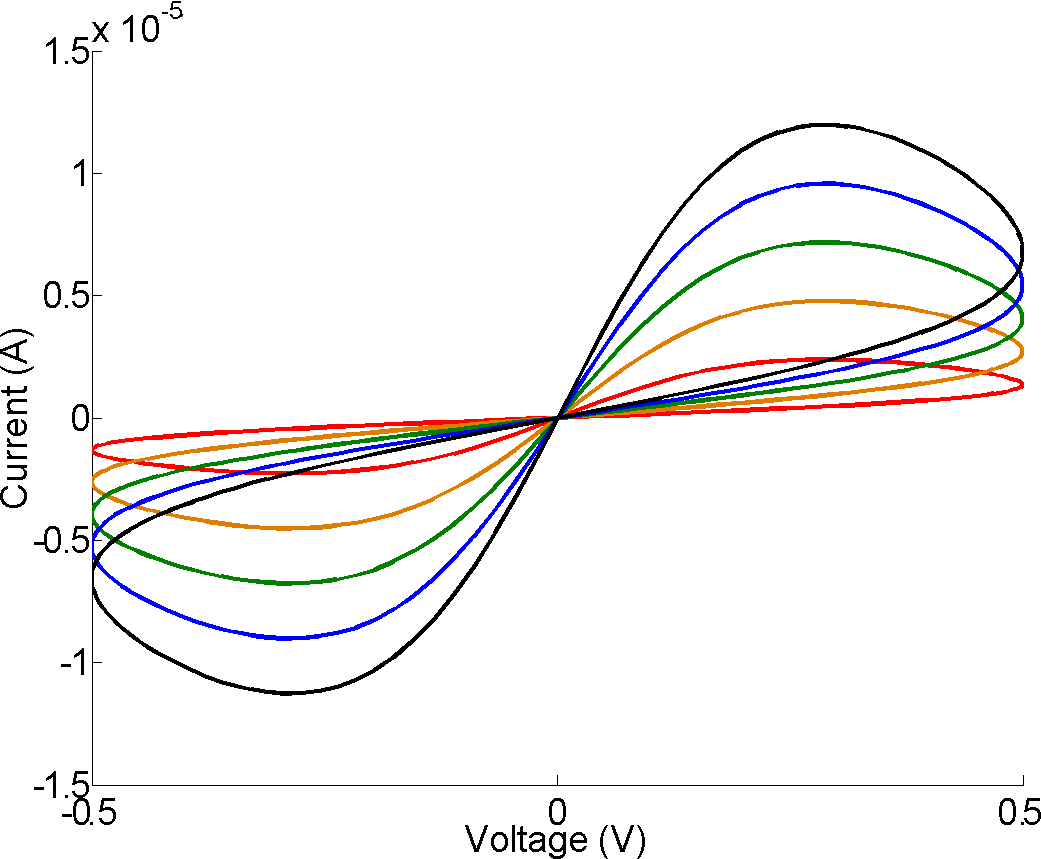}
  \caption{Ideal memristor V-I curves as calculated from the memory-conservation theory.}
  \label{fig.10}
\end{figure}

By adding in a contact resistance, which was measured from the actual devices at the start of the experiment, we obtain these predicted curves, which have the expected asymmetry, but are pinched.

\begin{figure}
\centering
\includegraphics[width=\textwidth]{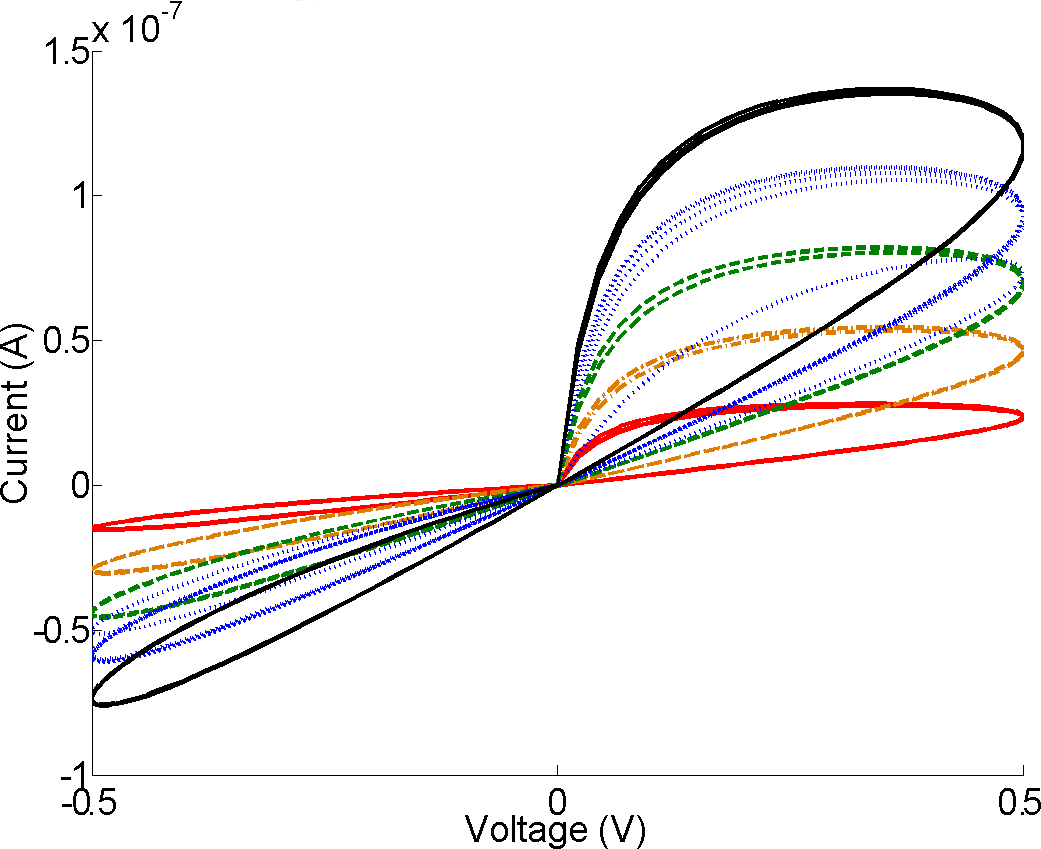}
  \caption{Pinched memristor curves obtained using the memory-conservation theory fit to the measured data (2 fitting parameters), with added measured contact resistance.}
  \label{fig.11}
\end{figure}

In this paper\cite{G1} (which I believe a draft version is up on arXiv), I then compared the measured hysteresis of the devices with the predicted hysteresis above and found that they were out by a constant ratio.

\begin{figure}
\centering
\includegraphics[width=\textwidth]{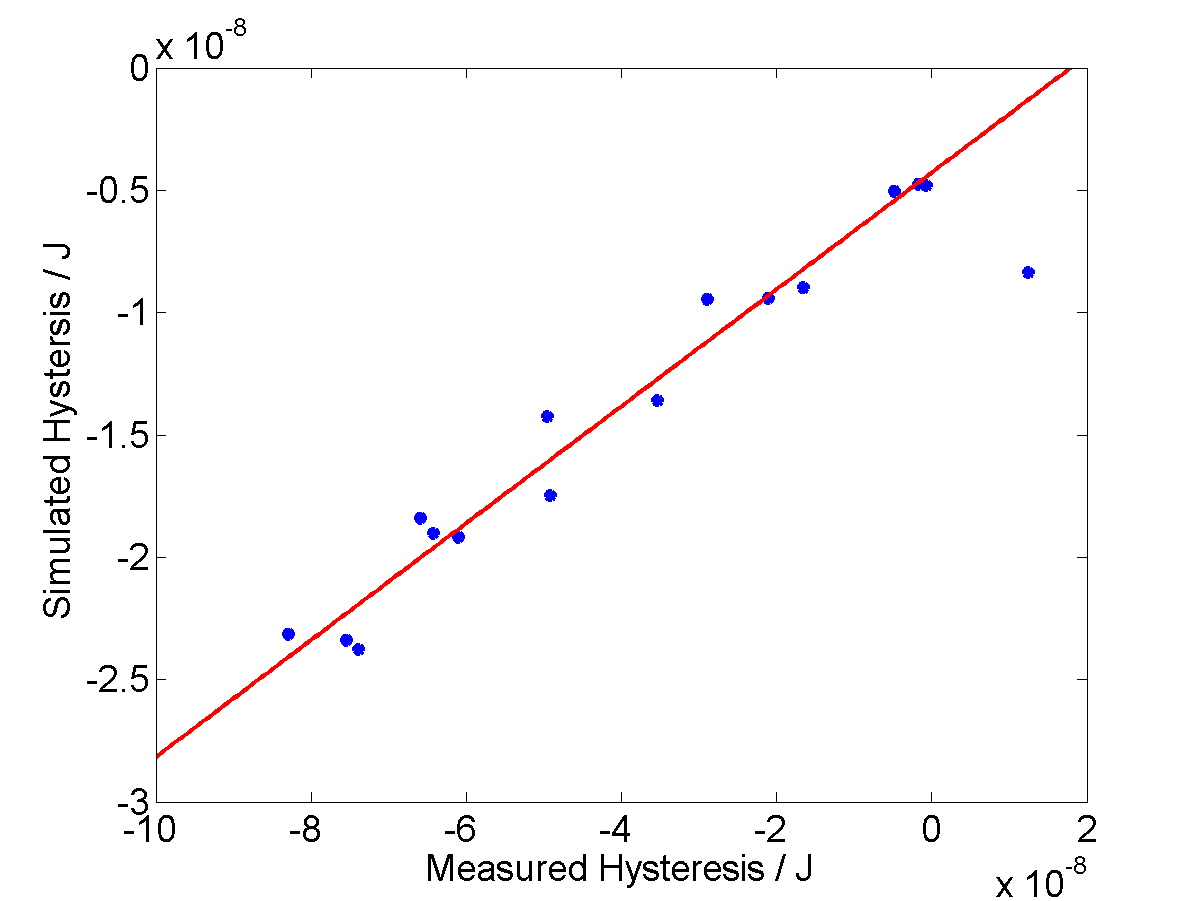}
  \caption{Comparison of the measured hysteresis from my devices with that predicted from the memory-conservation memristor theory with added contact resistance.}
  \label{fig.12}
\end{figure}

In that paper\cite{G1}, I merely suggested that as this relation was known, the theory could be used with a look up table to usefully predict device properties. From looking at the graphs in this letter, and considering R. S. William’s point, I now believe that the missing hysteresis is that of a capacitor, and that the capacitance increases with device size and I plan to add the capacitance into the model in that paper before publication.

\section{V-I curve summary}

So, I have shown that my devices exhibit the pinched memristor curve at high voltages and are thus accepted as being memristors, but that at low voltages, a capacitance can be seen close to 0V which prevents the memristor curves from crossing at 0. A memristor theory models the shape and asymmetry of the devices, but seems to require a capacitance in order to not cross at zero, this demonstrates that memristance is involved in the low voltage V-I curves, as well as in the high voltage V-I. These results suggest that my devices are non-ideal memristors, best modelled as memristor-capacitor circuits, and that their behaviour requires memristance to be properly explained. Now I shall turn to the description of my devices’ I-t behaviour. 

\begin{figure}
\centering
\includegraphics[width=\textwidth]{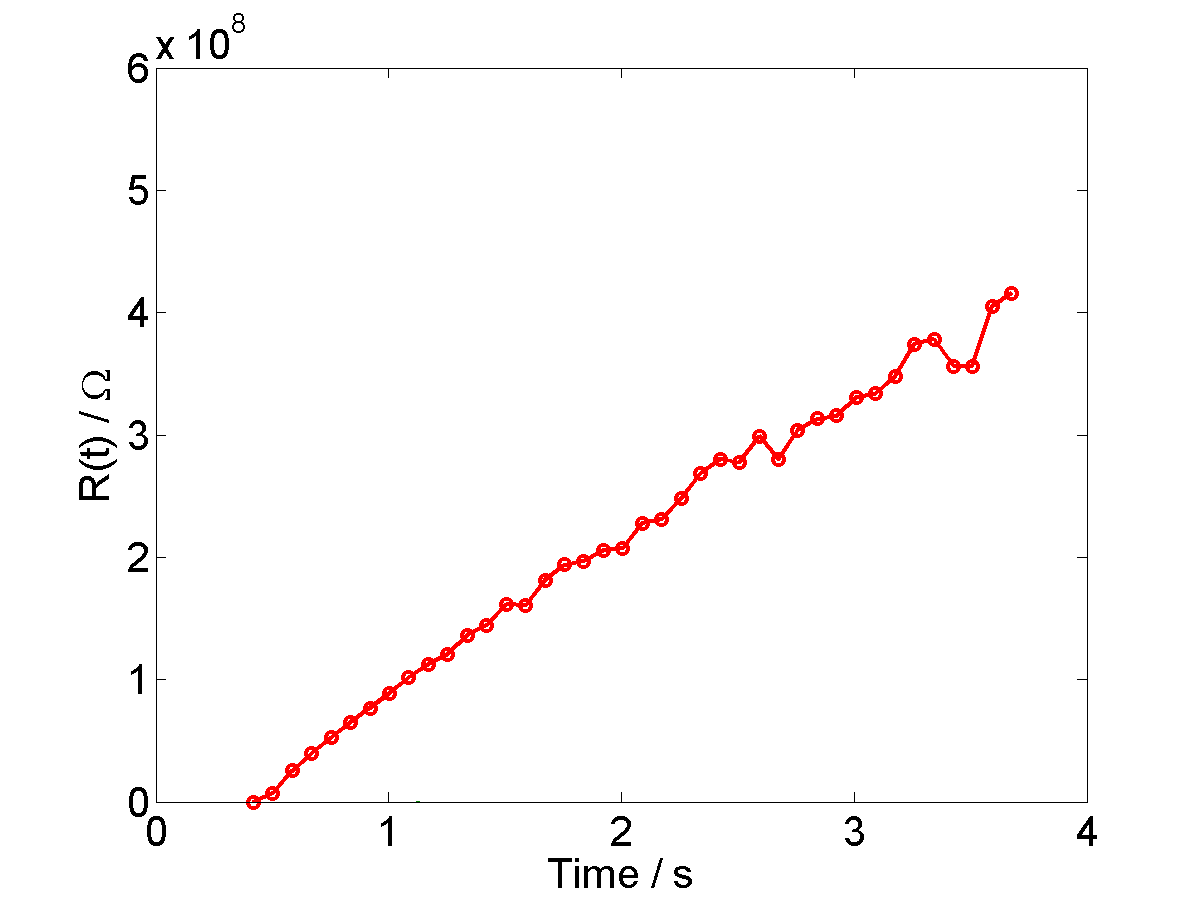}
  \caption{Rt}
  \label{fig.12a}
\end{figure}

\section{I-t curves under d.c. voltage}

\paragraph{Point 5: the memristor spikes are well fit by a memristor theory}

In \cite{SpC} I fitted both positive and negative memristor spikes to the memory conservation theory of memristance. For the positive spike, the summed square of residuals was 1.61x10$^{-17}$. For the negative spike, the summed square of residuals of 1.63x10$^{-17}$. For the exponential fit (to the short time spike), 2.43x10$^{-15}$. I got similar results with Erokhin et al’s 3-terminal memristor\cite{Berzina} (it was fit better by the memory conservation theory than an exponential). 

\begin{figure}
\centering
\includegraphics[width=\textwidth]{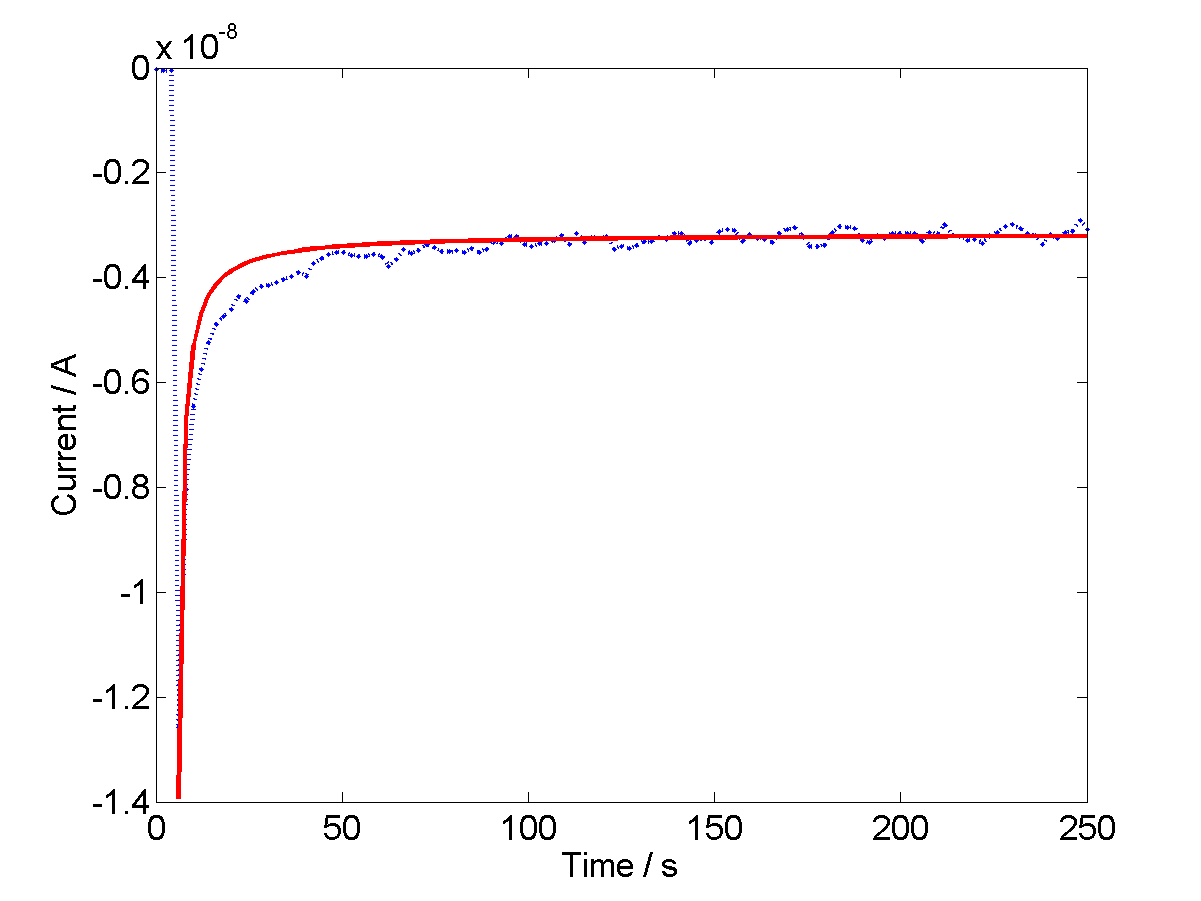}
  \caption{Negative current memristor spike (d.c. response) with memory conservation model fit. Taken from \cite{SpC}}
  \label{fig.13}
\end{figure}

\paragraph{Point 6. The d.c. responses (spikes) are related to the frequency effect in memristor hysteresis curves, which is one of memristor’s defining features.}

Finally, and most importantly, I will now demonstrate that these spikes are part of the definitions of memristance as they are behind the `frequency effect' described by L. O. Chua as a `fingerprint of the memristor'\cite{FG}. Doing experiments with several recording time steps at each voltage in a stepped triangular voltage waveform reveals spikes that decay, if one were to join up the currents measured the same number of dwell time-steps after an input voltage change, one draws a pinched I-V curve with shrinking hysteresis. As the third fingerprint of the memristor is that the ``pinched hysteresis loop should shrink to a single valued function when the frequency tends to infinity''\cite{FG}, the `frequency effect', this demonstrates that the frequency effect is directly related to the d.c. spikes, as I concluded in \cite{hystC}. 

The word `current transient' only means that the current is short-lived and does not actually define the cause (capacitance or memristance). When I first noticed these spikes I asked many memristor researchers if they had seen them in their devices, they had (although many ignored them), so I know that my devices are not unique in having this property (see my papers for when I found published memristor I-t curves). So, it is either the case that some previously observed `current transients' are memristive in form, or that the frequency effect is due to the capacitance found in non-ideal memristors (which does not fit with L. O. Chua’s equations, of course, as he modelled this effect in ideal memristors), or that R. S. Williams is correct, and my devices, and many other, possibly all, `memristors' are just leaky capacitors. 

\begin{figure}
\centering
\includegraphics[width=\textwidth]{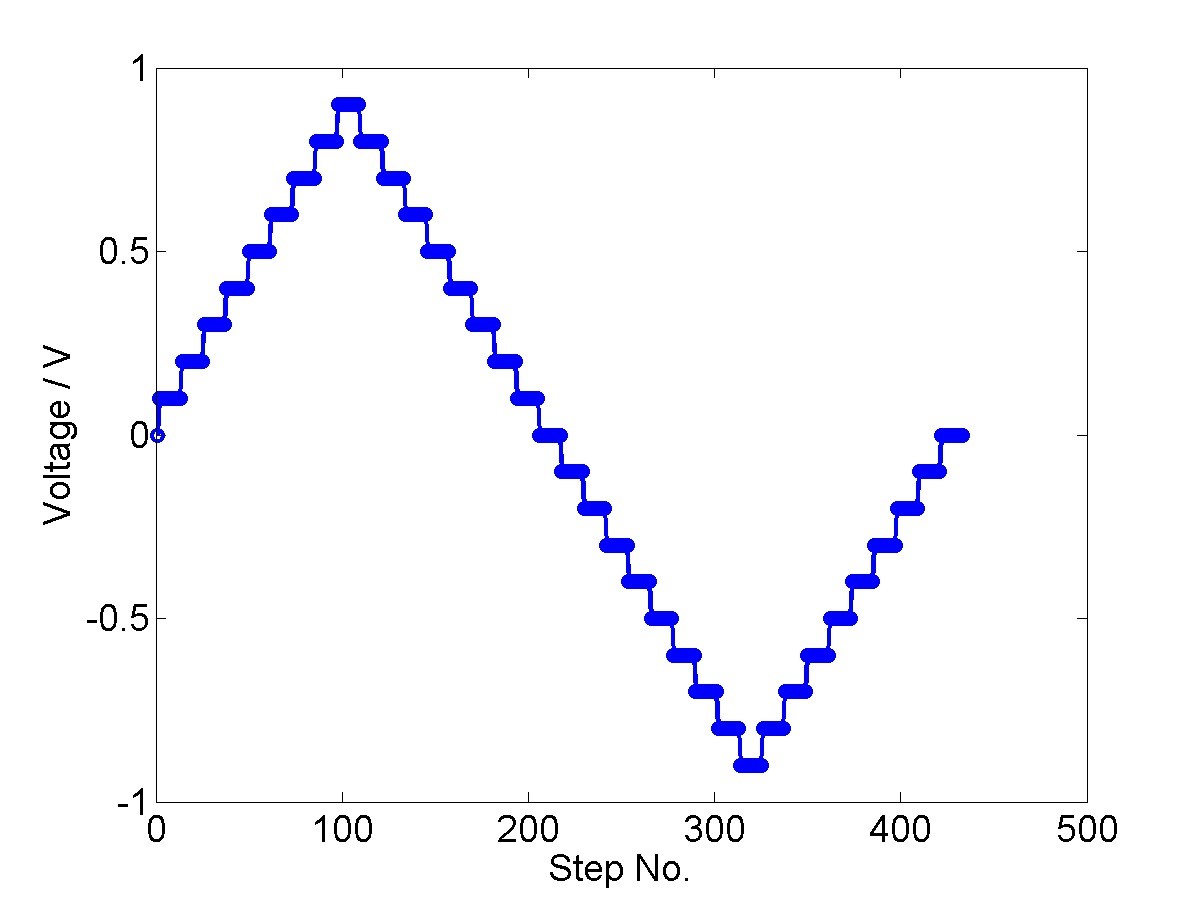}
  \caption{The V-t waveform used to investigate dc response and hysteresis. Taken from \cite{hystC}}
  \label{fig.14}
\end{figure}

\begin{figure}
\centering
\includegraphics[width=\textwidth]{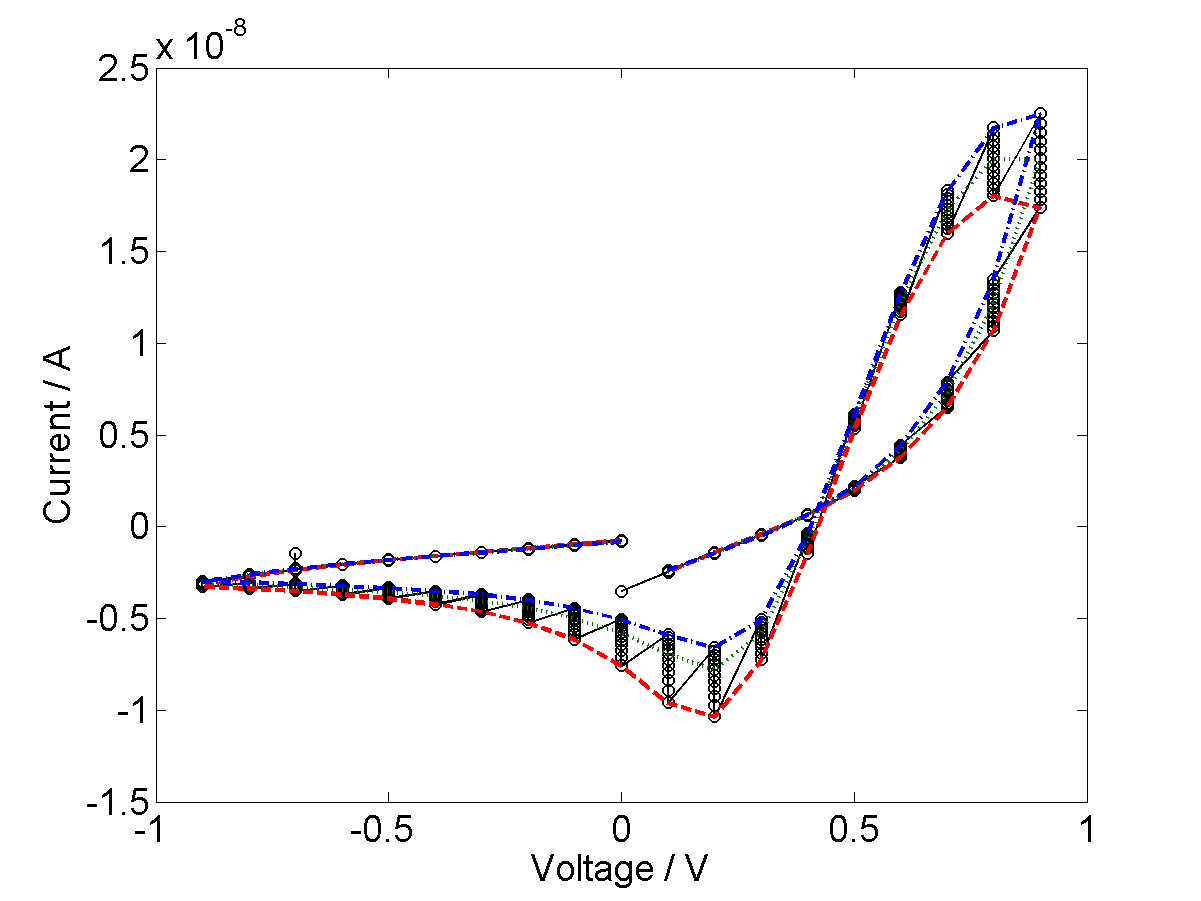}
  \caption{V-I curves for the PEO-PANI memristor under the V-t. The crossing point is offset in these devices. The first, 6$^{\mathrm{th}}$ and 12$^{\mathrm{th}}$ measurement points of each voltage have been joined together with red dashed, green dotted and blue dot-dashed lines respectively. The hysteresis loop shrinks with increased frequency. As each voltage step describes the d.c. response of the memristor, this demonstrates that the hysteresis effect is related to the d.c. spikes. Taken from \cite{hystC}}
  \label{fig.15}
\end{figure}

\begin{figure}
\centering
\includegraphics[width=\textwidth]{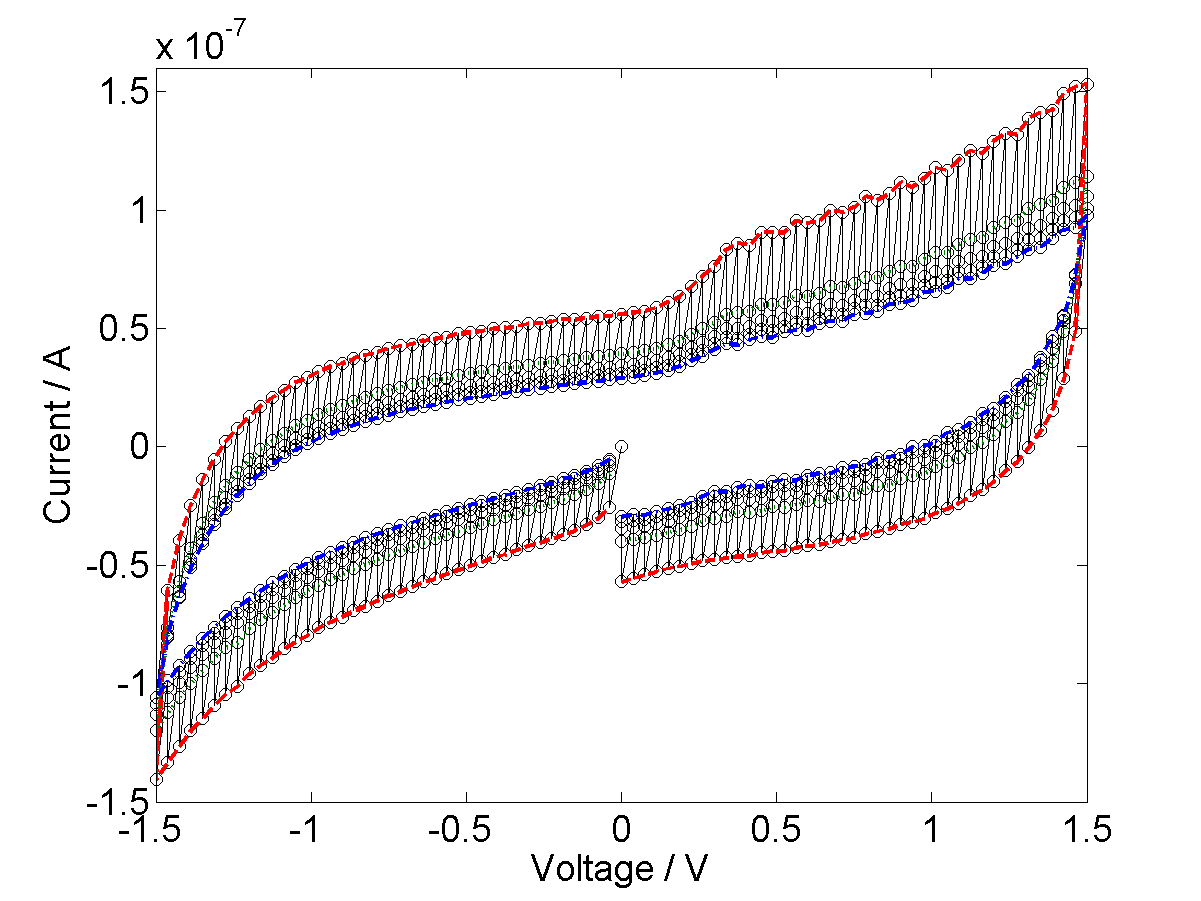}
  \caption{V-I curves for the Al-TiO$_2$-Al  memristor under small voltage V-t. The crossing point is offset in these devices. The first, 6$^{\mathrm{th}}$ and 12$^{\mathrm{th}}$ measurement points of each voltage have been joined together with red dashed, green dotted and blue dot-dashed lines respectively. The hysteresis loop shrinks with increased frequency. As each voltage step describes the d.c. response of the memristor, this demonstrates that the hysteresis effect is related to the d.c. spikes. Taken from \cite{hystC}}
  \label{fig.16}
\end{figure}

\begin{figure}
\centering
\includegraphics[width=\textwidth]{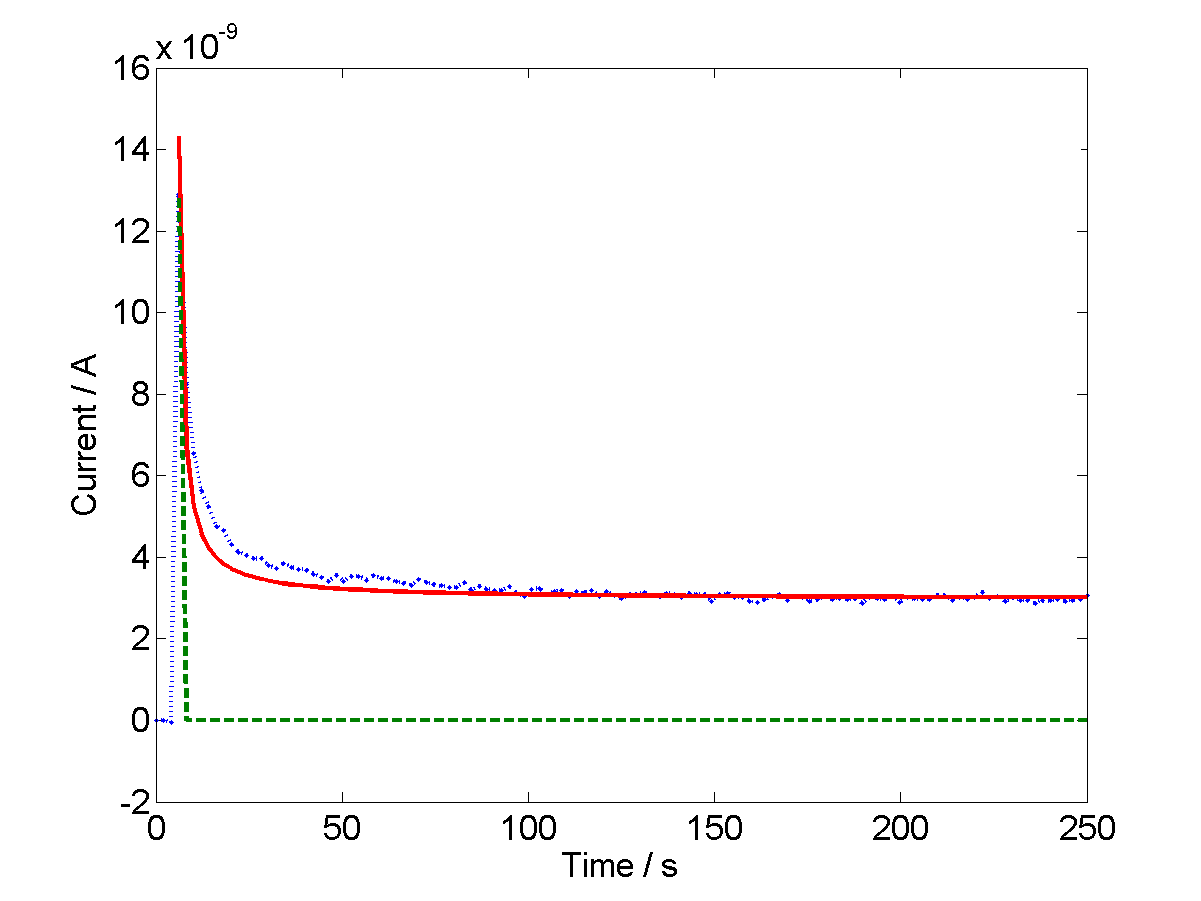}
  \caption{Positive memristor current spike with memory conservation theory fit. Taken from \cite{SpC}}
  \label{fig.17}
\end{figure}

\clearpage

\begin{acknowledgement}
Attendance at the New memory paradigms Faraday Discussion was partially supported by a Royal Society of Chemistry travel grant.

\end{acknowledgement}

\bibliography{ref}

\end{document}